\documentclass[%
 aip,
 amsmath,amssymb,
 reprint,%
]{revtex4-1}

\usepackage{graphicx}
\usepackage{dcolumn}
\usepackage{bm}

\usepackage[utf8]{inputenc}
\usepackage[T1]{fontenc}
\usepackage{mathptmx}
\usepackage{etoolbox}
\usepackage{float}
\usepackage{verbatim}
\usepackage{CJK}
\usepackage{indentfirst}
\usepackage{amsmath}
\usepackage{amsthm}
\usepackage{graphicx}
\usepackage{subfigure}
\usepackage{tabularx}
\usepackage{multirow}
\usepackage{algorithm}
\usepackage{algpseudocode}
\usepackage{xcolor}

\makeatletter
\def\@email#1#2{%
 \endgroup
 \patchcmd{\titleblock@produce}
  {\frontmatter@RRAPformat}
  {\frontmatter@RRAPformat{\produce@RRAP{*#1\href{mailto:#2}{#2}}}\frontmatter@RRAPformat}
  {}{}
}%
\makeatother
\begin{document}

\preprint{AIP/123-QED}

\title{A memory-based spatial evolutionary game with the dynamic interaction between learners and profiteers}
\author{Bin Pi}
 \affiliation{
 School of Mathematical Sciences, University of Electronic Science and Technology of China, Chengdu 611731, China.
 }
 
\author{Minyu Feng}
\affiliation{ 
College of Artificial Intelligence, Southwest University, Chongqing 400715, China.
}

 \author{Liang-Jian Deng}
 \email{liangjian.deng@uestc.edu.cn}
 \altaffiliation{Corresponding author}
 \affiliation{
 School of Mathematical Sciences, University of Electronic Science and Technology of China, Chengdu 611731, China.
 }


\date{\today}

\begin{abstract}

Spatial evolutionary games provide a valuable framework for elucidating the emergence and maintenance of cooperative behavior. However, most previous studies assume that individuals are profiteers and neglect to consider the effects of memory. To bridge this gap, in this paper, we propose a memory-based spatial evolutionary game with dynamic interaction between learners and profiteers. Specifically, there are two different categories of individuals in the network, including profiteers and learners with different strategy updating rules. Notably, there is a dynamic interaction between profiteers and learners, i.e., each individual has the transition probability between profiteers and learners, which is portrayed by a Markov process. Besides, the payoff of each individual is not only determined by a single round of the game but also depends on the memory mechanism of the individual. Extensive numerical simulations validate the theoretical analysis and uncover that dynamic interactions between profiteers and learners foster cooperation, memory mechanisms facilitate the emergence of cooperative behaviors among profiteers, and increasing the learning rate of learners promotes a rise in the number of cooperators. In addition, the robustness of the model is verified through simulations across various network sizes. Overall, this work contributes to a deeper understanding of the mechanisms driving the formation and evolution of cooperation.

\end{abstract}

\maketitle

\begin{quotation}

With the burgeoning development of artificial intelligence, reinforcement learning methods have become increasingly prevalent in exploring structured population behavior in evolutionary games. In light of the ubiquitous profit-seeking behavior observed in society and the inherent memory mechanisms of individuals, we propose a novel model in this paper, i.e., the memory-based spatial evolutionary game model with the dynamic interaction between learners and profiteers, where the memory mechanism is described by the memory length and the memory decay factor, and the dynamic interactions between learners and profiteers are modeled by a two-state homogeneous Markov chain. In addition, we conduct numerous simulations and analyses to verify the correctness of the theoretical derivations, the memory mechanism, and the dynamic interaction between two different categories of individuals on the impact of the frequency of cooperators, respectively, and study the emergence and evolution of cooperative behavior from a micro perspective.

\end{quotation}


\section{Introduction}\label{sec:introduction}

Cooperative behavior has been observed across various scales, ranging from microorganisms to complex animal societies, underscoring its ubiquity in the natural world. Scholars across disciplines, including sociologists \cite{simpson2015beyond}, psychologists \cite{henrich2021origins}, economists \cite{niyazbekova2023sustainable}, physicists \cite{guo2023third}, and mathematicians \cite{sun2023state}, etc. \cite{feng2023evolutionary, li2023open} have shown interest in understanding the origins and sustainability of cooperation. The network evolutionary game, as a combination of complex networks and evolutionary game theory, provides a practical framework for studying the emergence of cooperative behaviors in structured groups, where each node in a complex network represents an individual and the edges indicate the interactions between the individuals. Typical game models include the prisoner's dilemma game \cite{wang2022levy}, snowdrift game \cite{pi2022evolutionary2}, stag hunt game \cite{wang2013evolving} with two players, and the public goods game \cite{wang2022replicator} with multiple players. Besides, evolutionary games based on various structured populations have been widely proposed and studied, spanning square lattice networks with periodic boundaries \cite{flores2022cooperation, szabo2016evolutionary}, small-world networks \cite{lin2020evolutionary, chen2008promotion}, scale-free networks \cite{shen2024extortion, kleineberg2017metric}, temporal networks \cite{li2020evolution, sheng2023evolutionary}, and higher-order networks \cite{alvarez2021evolutionary, kumar2021evolution}. In addition to these, evolutionary games have recently gained attention and success in other areas as well \cite{capraro2024outcome}.

In recent years, researchers extensively explored the underlying drivers behind the spontaneous emergence and sustenance of cooperative behaviors within competitive environments, corroborating their findings through numerous simulation experiments. A seminal contribution is the five rules proposed by Nowak \cite{nowak2006five}. These rules encompass kin selection, direct reciprocity, indirect reciprocity, network reciprocity, and group selection, which elucidates diverse pathways to cooperation. Moreover, the mechanisms favoring cooperation include reputation mechanism \cite{xia2023reputation}, trust \cite{xie2024trust}, reward and punishment \cite{wang2014rewarding}, etc. \cite{wang2018exploiting, arefin2021imitation, zhang2023evolutionary}, which have also received extensive attention and investigation by scholars. In addition to these, many recent studies employ reinforcement learning methods to study the behavior of individuals in network evolutionary games, and Q-learning has especially become a dominant approach in these studies. For example, Yang et al. integrated Q-learning agents into the evolutionary prisoner's dilemma game on the square lattice with periodic boundary conditions and found that interaction state Q-learning promotes the emergence and evolution of cooperation \cite{yang2024interaction}. Shi and Rong delved into the dynamics of Q-learning and frequency adjusted Q-learning algorithms in multi-agent systems and revealed the intrinsic mechanisms of these algorithms from the perspective of evolutionary dynamics \cite{shi2022analysis}. Ding et al. explored the impact of Q-learning on cooperation by involving extortion and observed that Q-learning significantly boosts the cooperation level of the network \cite{ding2019q}. Therefore, amidst the rapid advancement of artificial intelligence, it is important to consider the influence of intelligent individuals equipped with learning on evolutionary dynamics.

In real systems, the memory of intelligent individuals significantly affects decision-making processes, and their actions are not limited to the current situation, but they also take past experiences into consideration. Some researchers discovered this phenomenon and achieved fruitful results in network evolutionary games. For example, the classical game tactics like generous-tit-for-tat (GTFT) \cite{nowak1992tit} and win-stay, lose-shift (WSLS) \cite{nowak1993strategy} proposed by Nowak are demonstrated to yield promising results in repeated prisoner's dilemma games. In addition, Pi et al. considered the memory mechanism and proposed two strategy-updating rules based on profiteers and conformists and found that the memory mechanism promotes cooperation over a large parameter area \cite{pi2022evolutionary}. Ma et al. examined the effect of working memory capacity, a crucial neural function, on cooperation in repeated prisoner's dilemma experiments and discovered that the level of cooperation was optimal when subjects remembered the first two rounds of information and that there was a sudden increase in the level of cooperation as memory capacity increased from none to minimal \cite{ma2021limited}. Lu et al. proposed a prisoner's dilemma game model with a memory effect on spatial lattices and observed that the memory effect could effectively change the cooperative behavior in the spatial prisoner's dilemma game \cite{lu2018role}. Therefore, the memory mechanism of the individual performs an indispensable role in the emergence of cooperative behaviors.

Profiteers commonly adhere to the Fermi rule, a strategy updating rule where individuals are more inclined to adopt the strategy of another individual with a higher payoff \cite{perc2010coevolutionary, yao2023inhibition, jusup2022social}. On the other hand, learners frequently employ Q-learning, a prevalent strategy updating rule where decisions are informed by previous learning experiences \cite{wang2024enhancing, zhu2023co, mcglohon2005learning}. However, most of the previous studies simplified the scenario by considering either profiteers or learners independently, while in practice, these two categories interact continuously, i.e., the neighbor of an individual consists of both profiteers and learners, and an individual is not maintaining a category all the time. Therefore, to bridge this gap, we consider the dynamic interaction between profiteers and learners in this paper, where each individual changes from a learner (profiteer) to a profiteer (learner) with a certain probability over time, which can be described as a two-state homogeneous discrete Markov chain. The different categories of individuals are mainly reflected in the different strategy updating rules they adopt. Concretely, the profiteer uses the classical Fermi rule, preferring to imitate the strategy of individuals with higher payoffs, whereas the learner employs Q-learning in reinforcement learning and decides which strategy to utilize by continuously learning from the past. Furthermore, as we mentioned before, memory acts as a crucial role in individuals' decision-making processes. Therefore, we introduce the memory mechanism into the evolution of the game. Specifically, the payoff of an individual is not solely dependent on a single game round but is a cumulative payoff, which is related to the memory length and the memory decay factor of the individual. This reflects the fact that an individual's memory is not infinite and a longer event has a smaller effect on the individual. Through our investigation of the memory-based snowdrift game with the dynamic interaction between profiteers and learners on regular square lattices with periodic boundary conditions and Watts-Strogatz small-world networks \cite{watts1998collective}, we find that the memory mechanism of individuals promotes the emergence and maintenance of cooperation among profiteers and the dynamic interaction between learners and profiteers enhances the cooperative behavior of the structured populations.

In the remainder of this paper, we first present the memory-based game model with the dynamic interaction between learners and profiteers in detail in Sec. \ref{Model}. Following that, in Sec. \ref{Simulation}, we show the simulation results and conduct thorough analyses. In the last section, we summarize the work and offer outlooks of this paper.

\vspace{-1.3\baselineskip}
\section{Model}
\label{Model}

In this section, we introduce the memory-based spatial evolutionary game with the dynamic interaction between learners and profiteers, which is described in terms of four aspects: (i) the game model, (ii) the memory mechanism, (iii) the dynamic interactions between learners and profiteers, and (iv) the stationary distribution of the number of learners and profiteers.

\vspace{-1.3\baselineskip}
\subsection{Game model}

In this study, we adopt the classical snowdrift game (SDG) for its generality, where the reward (R) for the interaction of two cooperative strategies is fixed to 1, the punishment (P) for the interaction of two defective strategies is set to 0, and for the interaction of cooperative and defective strategies, the cooperator receives a sucker's payoff (S) of $1 - r$ while the defector yields a temptation to defect (T) of $1 + r$. Hence, the payoff matrix of SDG is represented as follows:

\begin{equation}\label{Payoff matrix}
A = \left(                 
  \begin{array}{cc}  
    1 & 1 - r \\  
    1 + r & 0  \\  
  \end{array}
\right)                 
,
\end{equation}
where $r$ indicates the cost-to-benefit when both individuals are cooperators, and it takes a value ranging from 0 to 1, which is a flexible parameter.

\vspace{-1.3\baselineskip}
\subsection{Memory mechanism}

Next, we provide a detailed explanation of the memory mechanism proposed in this paper. Each individual in the system possesses a memory, i.e. they are capable of knowing their payoffs from the previous $M$ rounds of interactions with their neighbors. Hereby, $M$ denotes the length of the individual's memory, and the payoff in the past $M$ rounds has an impact on the individual's current payoff. However, it is acknowledged that the impact of past interactions diminishes over time. To account for this, we introduce a memory decay factor $\beta$, which characterizes the decreasing influence of past events on an individual's current payoff. Therefore, the payoff of individual $i$ at time $t$ based on the memory mechanism can be expressed as

\begin{equation}
\label{Payoff Calculation}
U_{i}^{t}=\sum_{k=t-M+1}^t{\beta ^{t-k}\Pi _{i}^{k}},
\end{equation}
where $\Pi _{i}^{k}$ represents the actual payoff of individual $i$ obtained from playing the snowdrift game with all neighbors in round $k$, and it can be calculated as

\begin{equation}
\Pi _{i}^{k}=\sum_{j\in \Omega_i}{s_{i}^{T}As_j},
\end{equation}
where $\Omega_i$ means the set consisting of all neighbors of individual $i$. We emphasize that when the evolutionary time is smaller than the memory length, i.e., $t < M$, then the individual's memory length is considered as $M = t$. Besides, $s_{x}$ is the strategy of individual $x$, where a unit vector $s_{x} = [0, 1]^T$ indicates a defective strategy and $s_{x} = [1, 0]^T$ denotes cooperation.

\vspace{-1.3\baselineskip}
\subsection{Dynamic interactions between learners and profiteers}

In reality, individuals exhibit diverse behaviors, with some driven primarily by profit-seeking motives (referred to as profiteers), while others rely on self-learning strategies (referred to as learners). Therefore, our proposed model incorporates the interaction between these two categories of individuals mentioned above. Furthermore, individuals are not fixed in their roles throughout the evolutionary game process, i.e., individuals cannot always be profiteers or learners, and they undergo mutual transitions between the two states. For this reason, we introduce a Markov process to capture this dynamic. Specifically, a profiteer (\textit{resp.} learner) changes to be a learner (\textit{resp.} profiteer) with the probability $q$ (\textit{resp.} $p$) at every moment, and the state transition matrix representing these probabilities is given by

\begin{equation}\label{Transition matrix}
B = \left(                 
  \begin{array}{cc}  
    1- p & p \\  
    q & 1 - q  \\  
  \end{array}
\right)
,
\end{equation}
where $p$ denotes the probability of an individual transitioning from a learner to a profiteer and $1 - p$ represents the probability of remaining a learner. Analogously, $q$ indicates the probability of an individual switching from a profiteer to a learner, while $1 - q$ denotes the probability of staying a profiteer. We highlight that there is no relationship between the transition probabilities $p$ and $q$. The only requirement is that they must satisfy the definition of probability, meaning both $p$ and $q$ must fall between 0 and 1.

For learners and profiteers, the key difference between them lies in their strategy updating rules. Profiteers tend to imitate the strategy of the neighbor with the highest payoff, employing the Fermi rule for strategy updating, i.e., at each round of evolution, a profiteer $i$ randomly selects an individual $j$ from its neighbors and adopts neighbor's strategy according to the following probability:

\begin{equation}
W_p(s_{i} \leftarrow s_{j})=\frac{1}{1+e^{(U_{i}-U_{j}) / \kappa}},
\end{equation}
where $s_i$ and $U_i$ signify the strategy expressed as unit vectors and payoff based on the memory mechanism of individual $i$, respectively. The parameter $\kappa$ means the noise factor, which is utilized to describe the irrational choices of individuals in the game. 

On the other hand, learners use a reinforcement learning algorithm known as Q-learning for strategy updates. Specifically, the Q-learning algorithm can be regarded as a Markov decision process, where the decision of an individual is only relevant to the current situation and is not influenced by past events, which can be represented by a tuple $(S, A, W_l, r)$. Hereby, $S = \{0C, 1C, 2C, \cdots, |\Omega_i|C\}$ (the number of cooperators among neighbors) and $A = \{Cooperate, Defect\}$ denote the state space and action space of individual $i$. $W_l: S\times A \rightarrow p$ is the state transition probability after adopting action $a\in A$ in state $s \in S$, and $r: S\times A \rightarrow U$ represents the reward of individual for performing action $a \in A$ in state $s \in S$, which can be obtained by Eq. \ref{Payoff Calculation}. Each learner owns a Q-table, which will be updated after each round of the game with all neighbors according to the following equation:

\begin{equation}
\label{q-learning}
Q_{S_t}^{A_t}(t+1)=Q_{S_t}^{A_t}(t)+\alpha [r_{S_t}^{A_t}(t+1)+\gamma \max_{a\in A} Q_{S_{t+1}}^{a}(t)-Q_{S_t}^{A_t}(t)],
\end{equation}
where $\alpha \in [0, 1]$ and $\gamma \in [0, 1]$ denote the learning rate and discount factor, respectively. A smaller $\gamma$ causes the individual to focus more on the immediate payoff, otherwise, the individual focuses more on past experiences. $Q_{S_t}^{A_t}(t)$ represents the utility obtained by the individual in state $S_t$ when taking action $A_t$ at time $t$. $r_{S_t}^{A_t}(t+1)$ indicates the immediate payoff gained by the individual in state $S_t$ at time $t+1$ when performing action $A_t$ at time $t$, which is determined by Eq. \ref{Payoff Calculation}. The term $\max_{a\in A} Q_{S_{t+1}}^{a}(t)$ signifies the maximum Q-value received by the individual in the future state $S_{t+1}$. Therefore, the learner learns to update the Q-table by constantly playing games with its neighbors during the evolutionary process to guide the updating of its strategy. In particular, to portray the random choice of individuals in real situations, which is a compromise between exploration and exploitation, we introduce the $\epsilon$-greedy algorithm. Specifically, at each decision point, exploration occurs with probability $\epsilon$, i.e., a strategy is randomly selected from the action set $A$ with a uniform probability distribution, and exploitation is performed with probability $1 - \epsilon$, i.e., the strategy with the highest Q-value in the current state is taken, and if there is more than one, then one is chosen randomly.

\vspace{-1.3\baselineskip}
\subsection{Stationary distribution of the number of learners and profiteers}

As we mentioned before, individuals are allowed to switch between profiteers and learners, with the corresponding transition matrix shown in Eq. \ref{Transition matrix}. The transition probability is independent of the initial moment, which can be viewed as a two-state homogeneous Markov chain $\{X_n, n = 0, 1, 2, \cdots\}$ with state space $E = \{Profiteer, Learner\}$, where $X_n$ denotes the category of each individual at time $n$. Moreover, for any $i, j\in E$, there always exists a positive integer $n_0$ such that $B_{i,j}^{n_0} > 0$, indicating the ergodicity of the Markov chain. By the ergodic theorem, we have:

\begin{equation}
\label{Stationary distribution}
\begin{cases}
	\left( \pi _1,\pi _2 \right) =\left( \pi _1,\pi _2 \right) B,\\
	\sum_{i=1}^2{\pi _i}=1,\\
\end{cases}
\end{equation}
where $(\pi_1, \pi_2)$ indicates the stationary distribution of the Markov chain $\{X_n, n = 0, 1, 2, \cdots\}$. Substituting Eq. \ref{Transition matrix} into Eq. \ref{Stationary distribution}, we obtain the solution $\pi_1 = q / (p + q)$ and $\pi_2 = p / (p + q)$, which implies that as evolutionary time extends indefinitely, the probability of an individual being a learner is $\pi_1$, while the probability of being a profiteer is $\pi_2$.

Furthermore, based on the fact that whether each individual is a profiteer or a learner is independent of the other individuals, i.e., the categories of individuals are independent of each other, we can yield the expected number of profiteers and learners in the network when the evolution stabilizes, as expressed in the following equation:

\begin{equation}
\label{Expectation of scale}
\begin{cases}
	E_{learner} = N \times \pi_1 = \frac{Nq}{p + q},\\
	E_{profiteer} = N \times \pi_2 = \frac{Np}{p + q},\\
\end{cases}
\end{equation}
where $N$ represents the size of the network.

\begin{center}
\begin{figure}[htbp]
\centering
\includegraphics[scale=0.33]{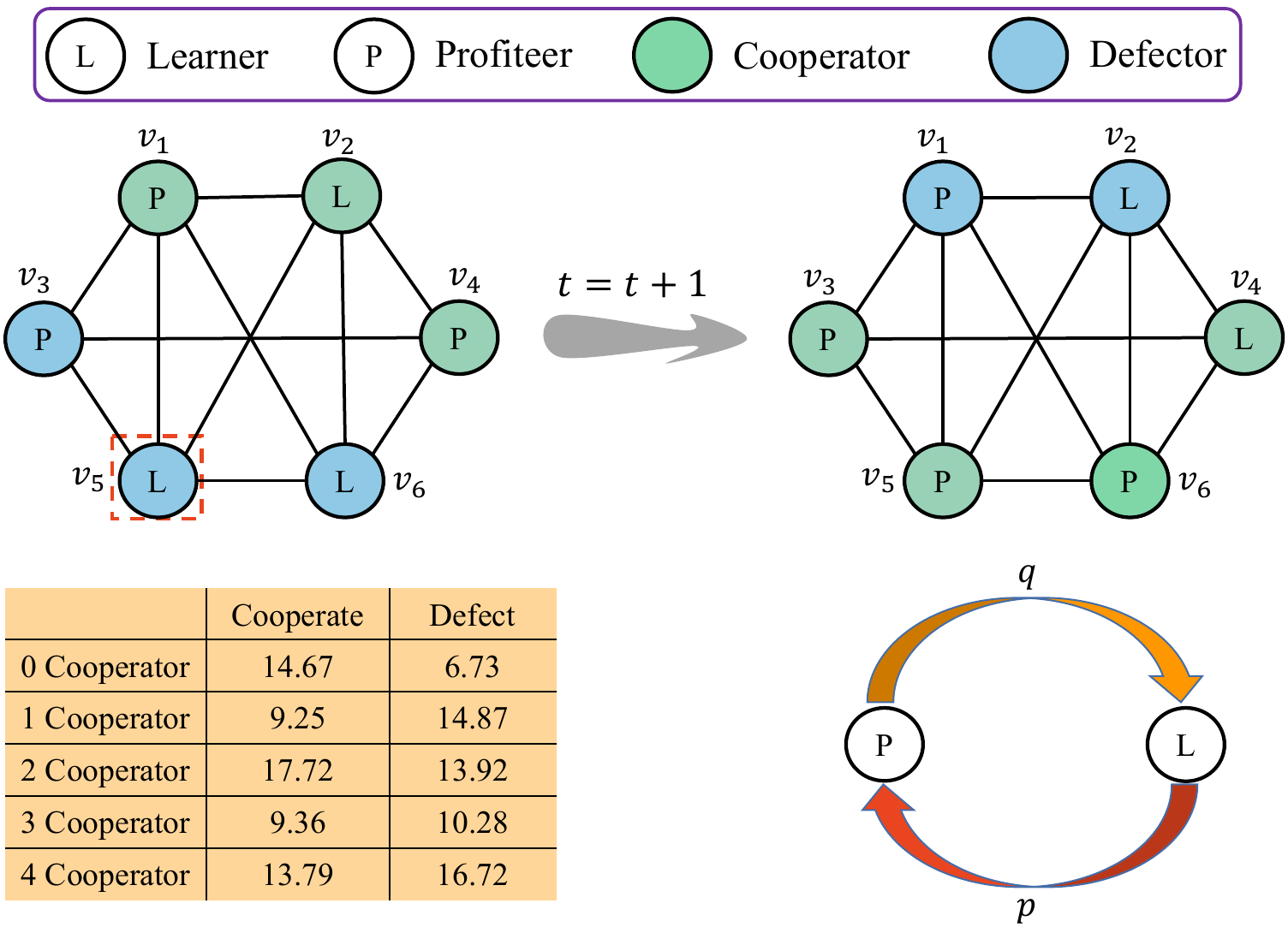}
\caption{\textbf{An illustration of the model.} In this figure, we provide an example of the proposed model. Profiteers and learners interact dynamically in the network, with different categories of individuals adopting different strategy updating rules to update their strategies. Besides, the category of each individual changes from learner (\textit{resp.} profiteer) to profiteer (\textit{resp.} learner) with the probability $p$ (\textit{resp.} $q$).}
\label{model illustration}
\end{figure}
\end{center}

To provide a clearer understanding of our model, we present an illustrative example of the model in Fig. \ref{model illustration}. Each individual possesses a probability of transitioning between being a profiteer and a learner, showcasing dynamic interactions between the two categories in the network. Profiteers and learners utilize Fermi rules and Q-learning to update their strategies, respectively. Notably, each learner maintains a Q-table. Taking individual $v_5$ highlighted within the red square as an example, it is surrounded by two cooperators and two defectors among its neighbors. According to the Q-table, we can get that the Q-value of adopting a cooperative strategy in the current state is 17.72, surpassing the Q-value (13.92) for choosing a defective strategy. Consequently, individual $v_5$ will opt for the cooperative strategy during exploitation at the next moment.

\vspace{-3.3\baselineskip}
\section{Simulation results and analysis}
\label{Simulation}

In this section, we aim to validate the impact of the proposed model on the evolution of cooperative behavior through numerical simulations and provide an analysis of the results. We conduct simulations on two categories of networks: (i) regular square lattice (SL) with periodic boundary conditions and von Neumann neighborhood with the network size of $N = 50 \times 50$; (ii) Watts-Strogatz small-world network (WS) with $N = 2500$, where each individual is initially connected to its 2 nearest neighbors on the left and right, and with a reconnection probability of 0.2. Initially, each individual is assigned to defect or cooperate with all its neighbors with a coin toss. Subsequently, individuals update their strategies using the reinforcement learning method called $\epsilon$-greedy Q-learning or the Fermi rule based on their categories, and the probability that a learner takes an exploration when updating the strategy is set to $\epsilon = 0.1$. We primarily focus on the evolution between profiteers and learners, the emergence of cooperative behaviors, and a microscopic view of the distribution of cooperators and defectors. In addition, we validate the robustness of the model by performing simulations on networks with different sizes. According to our simulations, the evolution of the cooperation frequency is stable after 1000 steps of iterations. Therefore, in all simulations, we average the last 500 of the entire 5000 time steps to obtain the result of one simulation. Additionally, to mitigate interference from other factors, we perform 20 independent simulations and take the average value of them to get the final outcome.

\subsection{Evolution and statistics of the number of profiteers and learners}

\begin{center}
\begin{figure}[htbp]
\centering
\subfigure[Evolutionary curves]{
\includegraphics[scale=0.3]{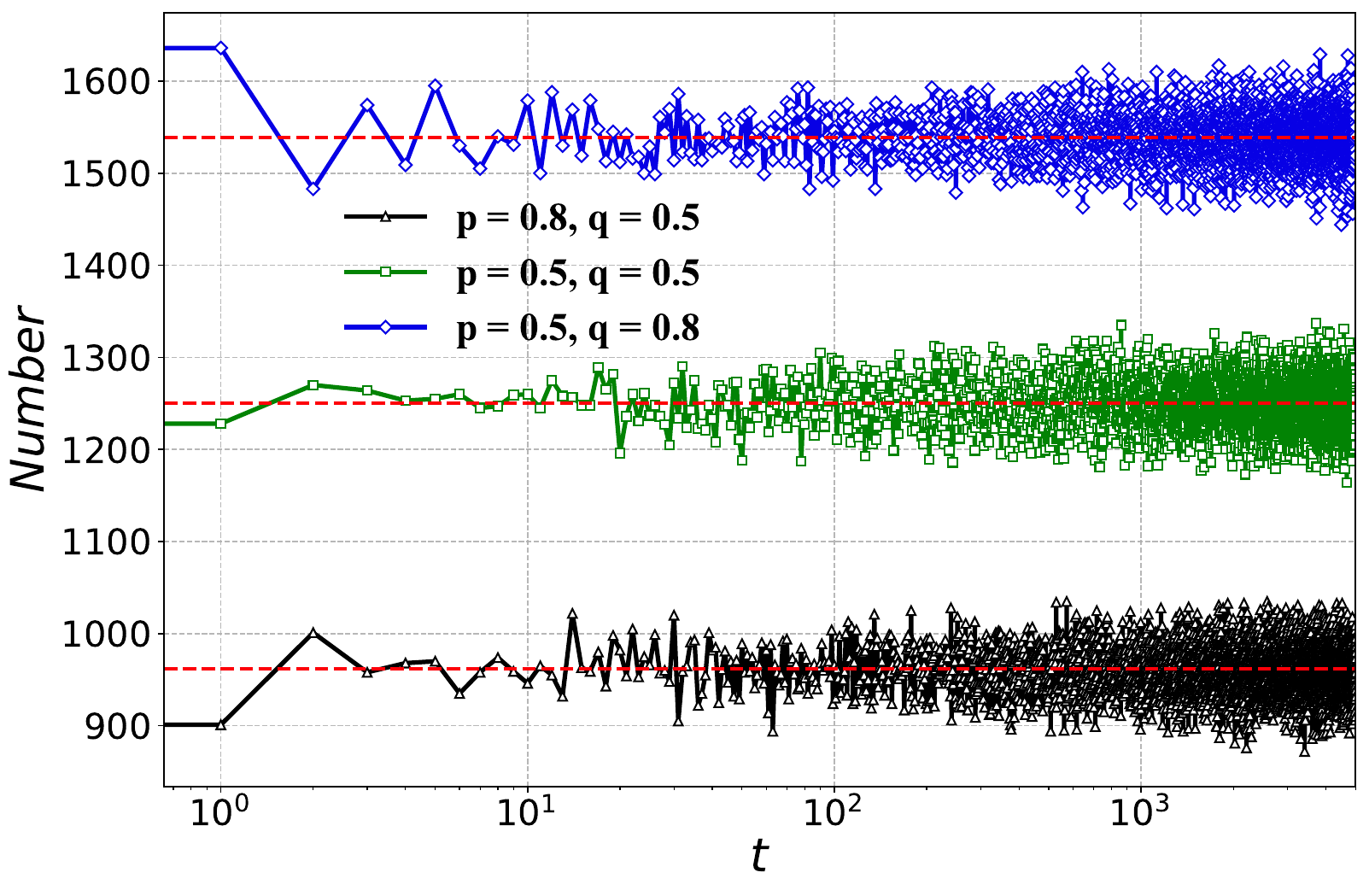}
\label{t_number}}
\subfigure[Statistical distributions]{
\includegraphics[scale=0.3]{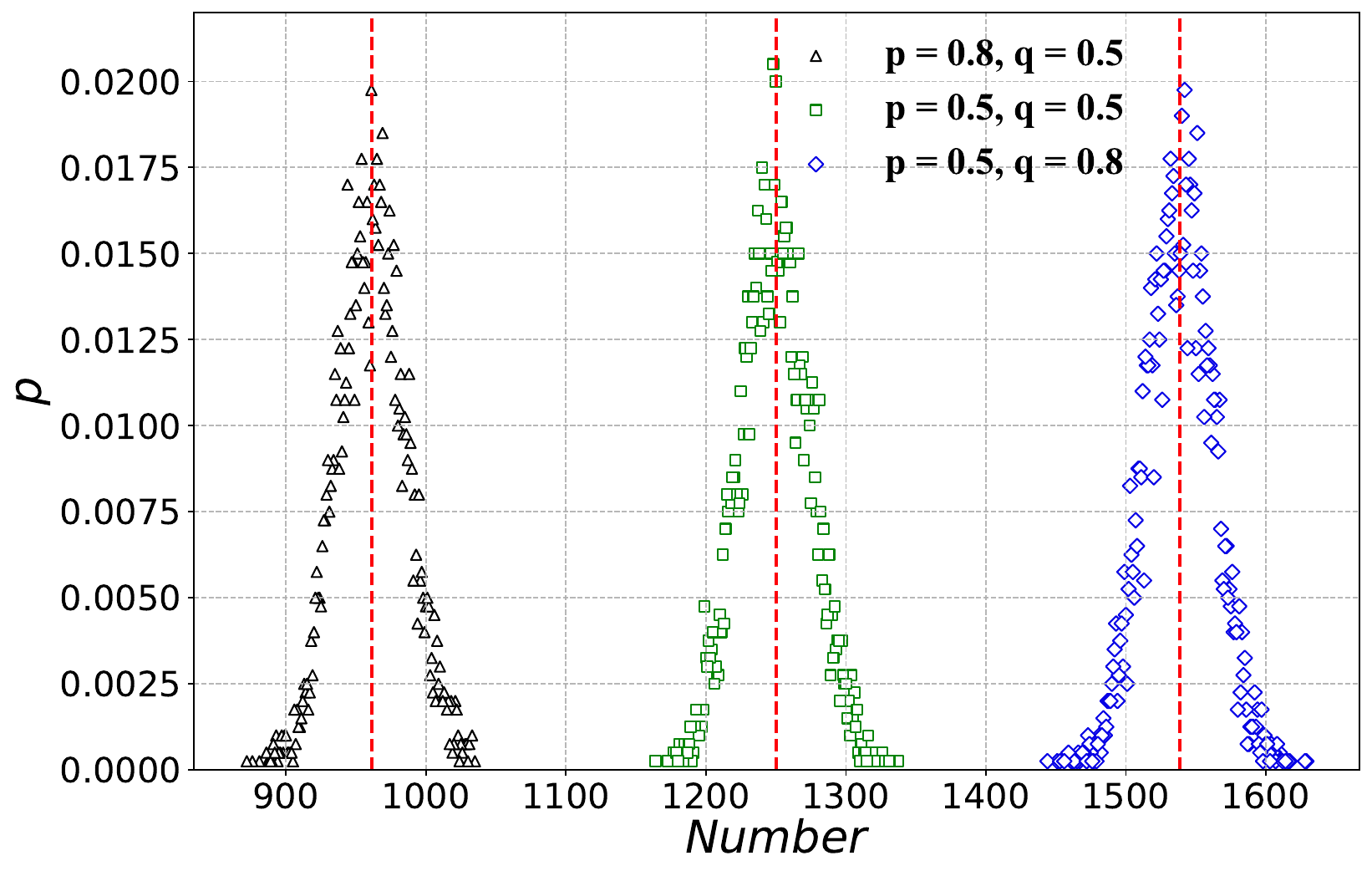}
\label{distribution_number}}
\caption{\textbf{Evolutionary curves and statistical distributions of the number of learners under different transition probabilities.} The green squares represent the results under $p = 0.5, q = 0.5$, while the blue diamonds and black triangles stand for the results under $p = 0.5, q = 0.8$ and $p = 0.8, q = 0.5$, respectively, with only one parameter different from green. The red line indicates the theoretical result derived from Eq. \ref{Expectation of scale}. We set the $x$-axis and $y$-axis as evolutionary time vs. number of learners for (a) evolutionary curves. In (b) statistical distributions, the $x$-axis represents the number of learners, while the $y$-axis shows the corresponding probability. It can be seen that the number of learners becomes stable around $t = 100$ and the numerical simulation results are in agreement with the theory.}
\label{number}
\end{figure}
\end{center}

In order to verify the theory of dynamic interaction between learners and profiteers proposed in this paper, we first plot the evolutionary curves of the number of learners over time under different transition probabilities of $p$ and $q$, and the result is shown in Fig. \ref{t_number}, where the red straight line indicates the theoretical value, which is calculated according to Eq. \ref{Expectation of scale}. We do not plot the evolution of profiteers since it can be obtained by subtracting the number of learners from the size of the network. From Fig. \ref{t_number}, we can see that the number of learners in all three cases evolves over time and stabilizes around $t = 100$, subsequently fluctuating around a certain value, which closely aligns with the one marked by the red theoretical straight line. Additionally, we present the theoretical and simulated values of the number of learners and profiteers, along with the relative error between them in Tab. \ref{statistics of plot} in the form of data for the three different scenarios. This provides a more intuitive insight into the gap between theory and simulation. The theoretical value is calculated using Eq. \ref{Expectation of scale}, while the simulated value is obtained by averaging the last 1000 steps of the evolution curve in Fig. \ref{t_number}. The relative error is determined by $e = |x - x^*| / x^*$, where $x^*$ and $x$ denote the theoretical and simulated values, respectively. We find that the theoretical and simulated results for all three situations are very close to each other, and the maximum relative error is only 0.096\%, which confirms the correctness of our theoretical derivation. Furthermore, by comparing the number of learners and profiteers at the stable time under the three cases, we observe that increasing $p$ leads to an increase in the number of profiteers, while a larger $q$ results in more learners in the network, which is consistent with the results demonstrated in Fig. \ref{t_number}.

\vspace{-1.8\baselineskip}
Subsequently, we record the number of learners in the last 4000 steps in each scenario and plot the probability distribution of the number of learners under three sets of transition probabilities according to the law of large numbers by considering the frequency as probability. The result is shown in Fig. \ref{distribution_number}, where the red straight line represents the theoretical values. Similarly, we present some numerical characteristics for each of the three distributions in Tab. \ref{statistics of distribution}, including standard deviation, skewness, and kurtosis, where the standard deviation is calculated using the $std()$ method in the $numpy$ library in Python, while the skewness and kurtosis are obtained employing the $skew()$ and $kurt()$ methods in the $pandas$ library, respectively. We observe that all three distributions approximately follow a normal distribution and that the span of each distribution is relatively small, explaining the small standard deviation of each distribution, which is consistent with the result in Tab. \ref{statistics of distribution}. Moreover, it can be clearly seen that the red theoretical straight line exactly passes through the tip of each distribution, which also verifies the correctness of our theory in another way. The skewness of all three statistical distributions of the number of learners exhibited in Tab. \ref{statistics of distribution} is less than 0. It indicates that all three distributions have negative skewness, i.e., the left skewness, where there are fewer data located on the left side of the mean than on the right side of the mean, whereas profiteers are the reverse. The kurtosis of learners under both $p = 0.8, q = 0.5$ and $p = 0.5, q = 0.5$ is less than 0, which means that the overall data distribution is relatively flat compared to the normal distribution and is platykurtic, while the kurtosis under $p = 0.5, q = 0.8$ is greater than 0, which implies that the overall data distribution is relatively steep compared to the normal distribution and is leptokurtic. It is worth noting that the standard deviation and kurtosis are the same for both profiteers and learners, while the skewness is just the opposite. This discrepancy arises because the network contains only two categories of individuals, profiteers and learners, and the standard deviation and kurtosis are essentially even-ordered central moments, whereas the skewness is an odd-ordered central moment.

\begin{center}
\begin{table*}[htbp]
\renewcommand{\arraystretch}{1.8}
\setlength{\tabcolsep}{8pt}
\begin{center}
\caption{The comparison of theoretical and simulated results for the number of learners and profiteers}
\begin{tabular}{c|cc|cc|cc}
\hline
\multirow{2}{*}{Results} & \multicolumn{2}{c|}{$p = 0.8, q = 0.5$} & \multicolumn{2}{c|}{$p = 0.5, q = 0.5$} & \multicolumn{2}{c}{$p = 0.5, q = 0.8$} \\ \cline{2-7}
                         & Learner    & Profiteer    & Learner    & Profiteer    & Learner    & Profiteer  \\ \hline
Theoretical values      & 961.538    & 1538.462    & 1250   & 1250      & 1538.462      & 961.538      \\
Simulated values       & 961.379    & 1538.621    & 1248.799   & 1251.201      & 1538.403     & 961.597   \\
Relative error           & 0.017\%      & 0.010\%      & 0.096\%     & 0.096\%      &  0.004\%     & 0.006\%   \\
\hline
\end{tabular}
\label{statistics of plot}
\end{center}
\end{table*}
\end{center}

\begin{center}
\begin{table*}[htbp]
\renewcommand{\arraystretch}{1.8}
\setlength{\tabcolsep}{8pt}
\begin{center}
\caption{The numerical characteristics for the distribution of the number of learners and profiteers}
\begin{tabular}{c|cc|cc|cc}
\hline
\multirow{2}{*}{Results} & \multicolumn{2}{c|}{$p = 0.8, q = 0.5$} & \multicolumn{2}{c|}{$p = 0.5, q = 0.5$} & \multicolumn{2}{c}{$p = 0.5, q = 0.8$} \\ \cline{2-7}
                   & Learner    & Profiteer    & Learner    & Profiteer    & Learner    & Profiteer  \\ \hline
Standard deviation      & 24.148    & 24.148    & 25.905   & 25.905      & 24.999      & 24.999      \\
Skewness       & -0.038    & 0.038    & -0.021   & 0.021      & -0.081     & 0.081   \\
Kurtosis         & -0.144      & -0.144     & -0.223     & -0.223      &  0.451     & 0.451  \\
\hline
\end{tabular}
\label{statistics of distribution}
\end{center}
\end{table*}
\end{center}

\vspace{-3.3\baselineskip}
\subsection{Emergence and evolution of cooperative behavior}

\begin{center}
\begin{figure}[htbp]
\centering
\subfigure[SL network]{
\includegraphics[scale=0.27]{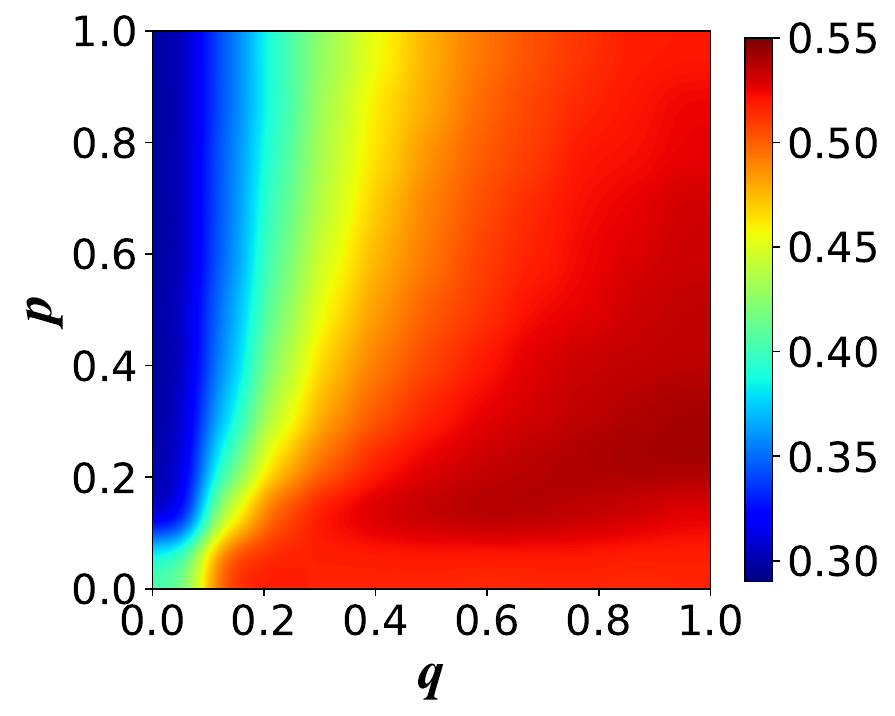}
\label{p_q_SL}}
\subfigure[WS network]{
\includegraphics[scale=0.27]{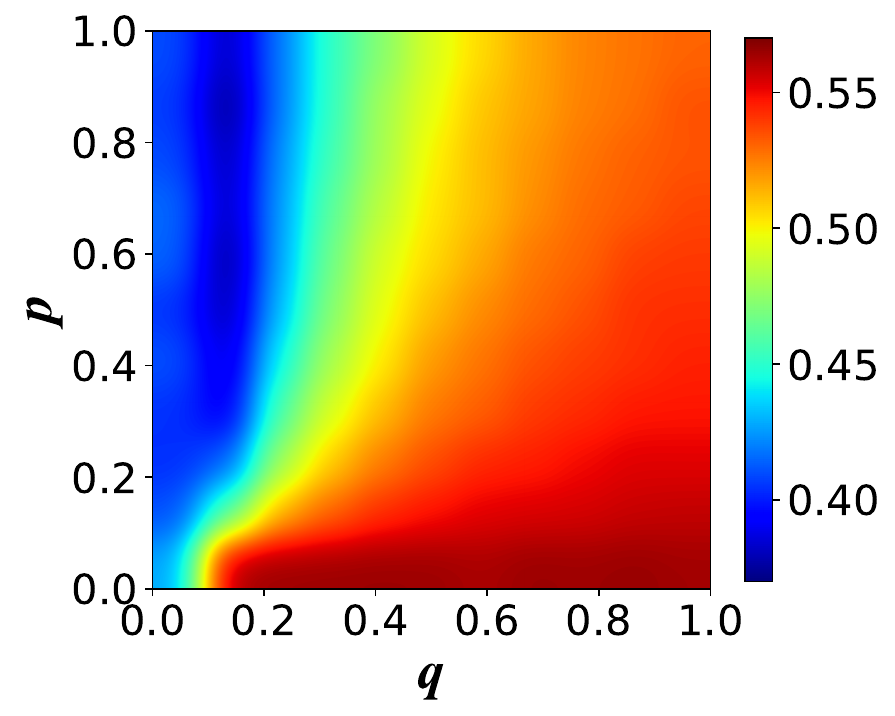}
\label{p_q_WS}}
\caption{\textbf{Heat maps of cooperation ratio regarding transition probabilities $p$ and $q$.} This figure elucidates the impact of transition probabilities $p$ ($y$-axis) and $q$ ($x$-axis) between profiteers and learners on cooperative behavior on the (a) SL and (b) WS networks. The parameters for the cost-to-benefit of SDG, memory decay factor, and memory length on both networks are fixed to $r = 0.6$, $\beta = 0.5$, and $M = 5$ respectively. These heat maps provide a comprehensive visualization of how varying $p$ and $q$ affect cooperative dynamics within different network topologies. We observe that the inclusion of learners boosts the frequency of cooperators compared to pure profiteers.}
\label{p_q_f}
\end{figure}
\end{center}

In this section, we investigate the effect of some model parameters on the emergence and evolution of cooperative behaviors. Primarily, we present the heat map of the cooperation ratio with respect to the transition probabilities $p$ and $q$ between profiteers and learners, as shown in Fig. \ref{p_q_f}. The $x$-axis is set as $q$ with the range [0, 1], which denotes the transition probability from profiteers to learners, and the $y$-axis is set as $p$ with the same range, which means the transition probability from learners to profiteers. Both networks exhibit that when $p$ is fixed, increasing the transition probability $q$ leads to a rise in the proportion of cooperators, i.e., the incorporation of learners promotes the emergence of cooperative behavior compared to a population of pure profiteers. This insight offers a fresh perspective to explain the widespread cooperative behavior in real-world scenarios. Although many individuals strive to behave as profiteers by imitating the strategies of individuals with high payoffs, there are still learners who are self-learning based on their past experiences, and the existence of learners is exactly the reason that further enhances the proportion of cooperators in the population.

\begin{center}
\begin{figure}[htbp]
\centering
\subfigure[SL network]{
\includegraphics[scale=0.27]{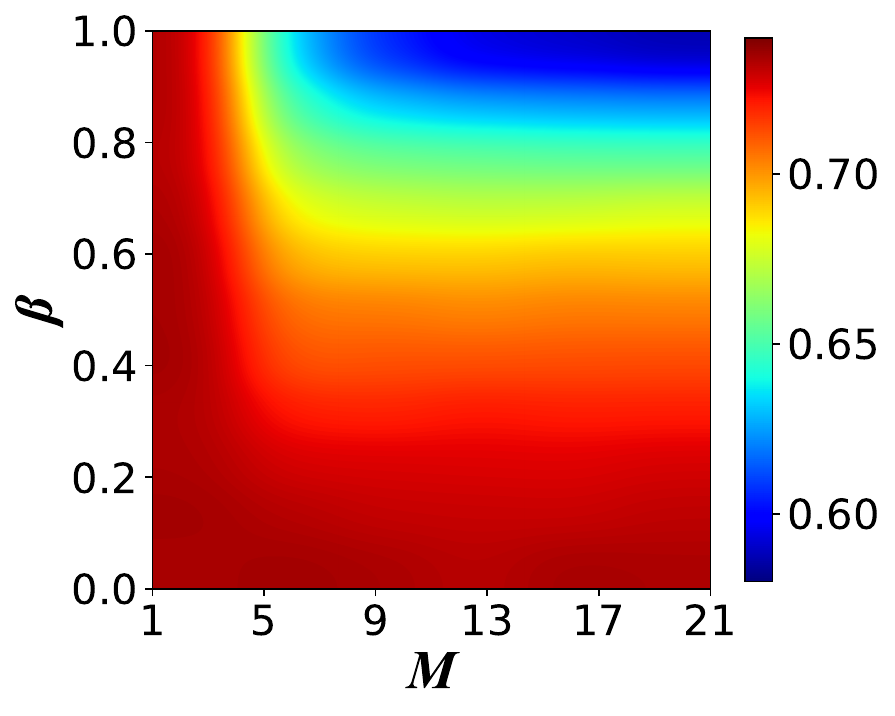}
\label{mdc_T_SL_mix}}
\subfigure[WS network]{
\includegraphics[scale=0.27]{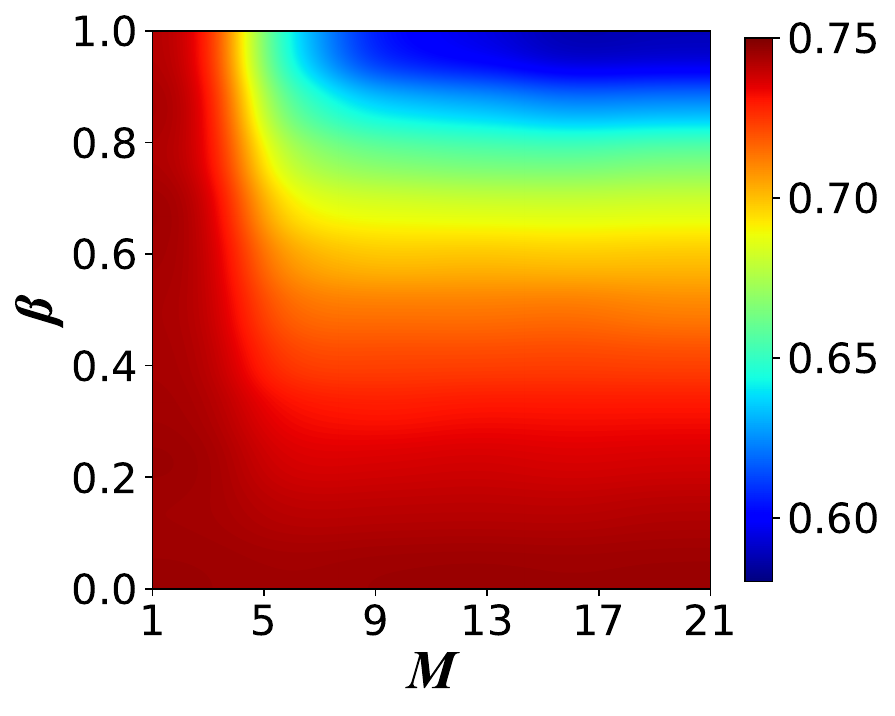}
\label{mdc_T_WS_mix}}
\caption{\textbf{Heat maps of cooperation frequency regarding memory length and memory decay factor under the coexistence of profiteers and learners.} In this figure, we illustrate the influence of memory decay factor $\beta$ and memory length $M$ on the frequency of cooperators on the SL (in panel (a)) and WS (in panel (b)) networks under the coexistence of profiteers and learners. We set the transition probabilities between learners and profiteers to $p = q = 0.5$, ensuring a balanced presence of learners and profiteers in the network. Additionally, the payoff parameter of the SDG is set to $r = 0.3$. These heat maps offer insights into how different combinations of memory parameters influence cooperative behavior in networks with varied structures. Both networks show that increasing memory length and memory decay factor inhibit cooperative behavior.}
\label{mdc_T_f_mix}
\end{figure}
\end{center}

\vspace{-2.8\baselineskip}
Subsequently, we delve into the impact of the memory mechanisms proposed in this paper on the evolution of cooperative behavior. We set the transition probability between profiteers and learners as $p = q = 0.5$, thus we can get that the profiteers and learners in the network are uniformly mixed according to Eq. \ref{Expectation of scale}. In this case, we plot the heat map of the cooperation ratio concerning the memory length and the memory decay factor, and the results of SL and WS networks are respectively shown in Figs. \ref{mdc_T_SL_mix} and \ref{mdc_T_WS_mix}, from which we can obtain that increasing both the memory length $M$ and the memory decay factor $\beta$ causes a decrease in the number of cooperators for both SL and WS networks.

\begin{center}
\begin{figure}[htbp]
\centering
\subfigure[SL network]{
\includegraphics[scale=0.27]{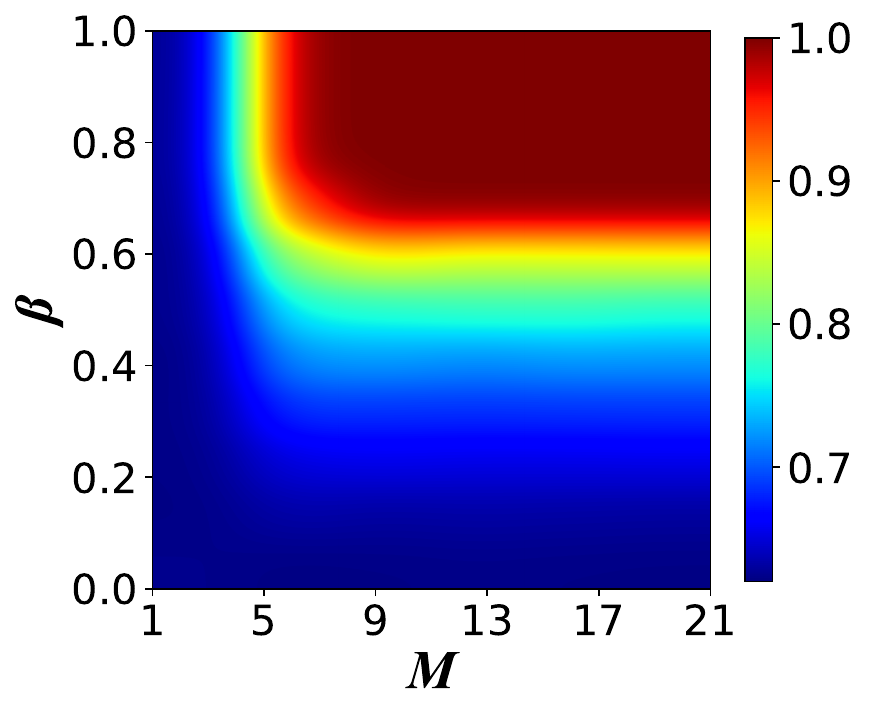}
\label{mdc_T_SL_profiteer}}
\subfigure[WS network]{
\includegraphics[scale=0.27]{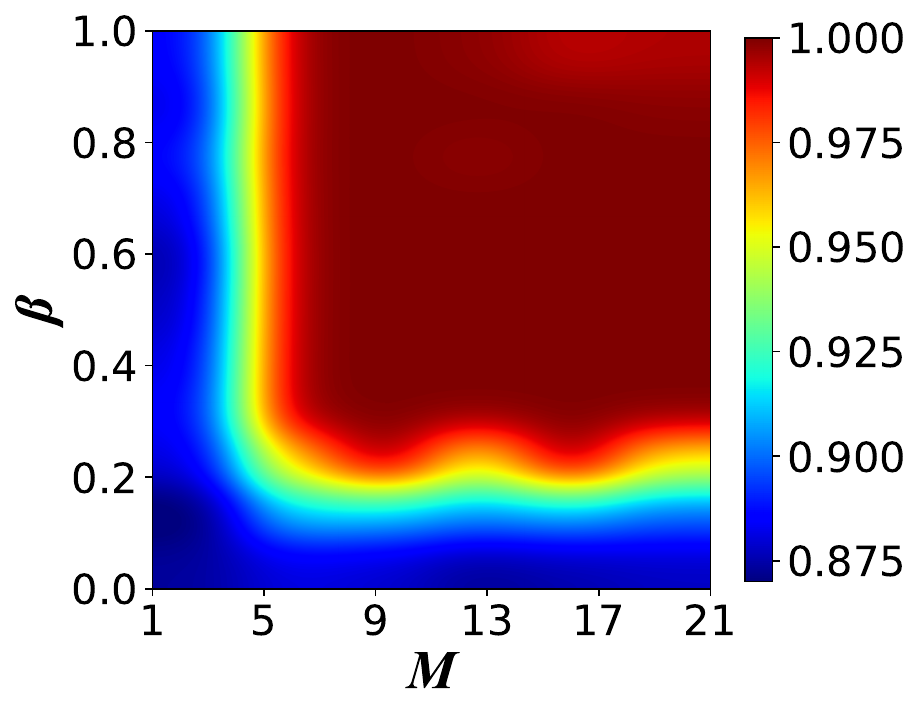}
\label{mdc_T_WS_profiteer}}
\caption{\textbf{Heat maps of cooperation frequency regarding memory length and memory decay factor in the case of pure profiteers.} This figure presents the influence of memory decay factor $\beta$ ($y$-axis) and memory length $M$ ($x$-axis) on the cooperative behavior on the SL (in subplot (a)) and WS (in subplot (b)) networks under conditions of pure profiteers. Hereby, transition probabilities between learners and profiteers are fixed to $p = 1$ and $q = 0$, resulting in all the individuals in the network being profiteers, with no learners existing. Furthermore, the payoff parameter of SDG is set to $r = 0.3$. These heat maps provide insights into how memory parameters influence cooperative behavior in networks dominated by profiteers. It is evident that memory mechanisms favor the survival of cooperators in groups of pure profiteers.}
\label{mdc_T_f_profiteer}
\end{figure}
\end{center}

\vspace{-2.3\baselineskip}
In addition, we demonstrate the changes in the proportion of cooperators concerning the memory mechanism for the pure profiteer scenario in Fig. \ref{mdc_T_f_profiteer}, where the probability of converting from learner to profiteer is set to $p = 1$. In contrast, the probability of changing from profiteer to learner is $q = 0$. All other parameters are the same as in the numerical simulation in Fig. \ref{mdc_T_f_mix}. It is obvious that the memory mechanism greatly facilitates the emergence of cooperative behavior in the case of pure profiteers. Notably, pure cooperators even appear on both networks when the memory length and memory decay factor are relatively large. Furthermore, upon comparing Figs. \ref{mdc_T_SL_profiteer} and \ref{mdc_T_WS_profiteer}, we can observe that the area of the pure cooperators' region of WS shown in Fig. \ref{mdc_T_WS_profiteer} is significantly larger than that of SL presented in Fig. \ref{mdc_T_SL_profiteer}, which suggests that the WS network is more favorable to the survival of cooperators than the SL network. Therefore, we can conclude that while the memory mechanism facilitates the emergence of cooperative behavior in profiteers, it inhibits the evolution of cooperation in learners.

\begin{center}
\begin{figure}[htbp]
\centering
\subfigure[SL network]{
\includegraphics[scale=0.27]{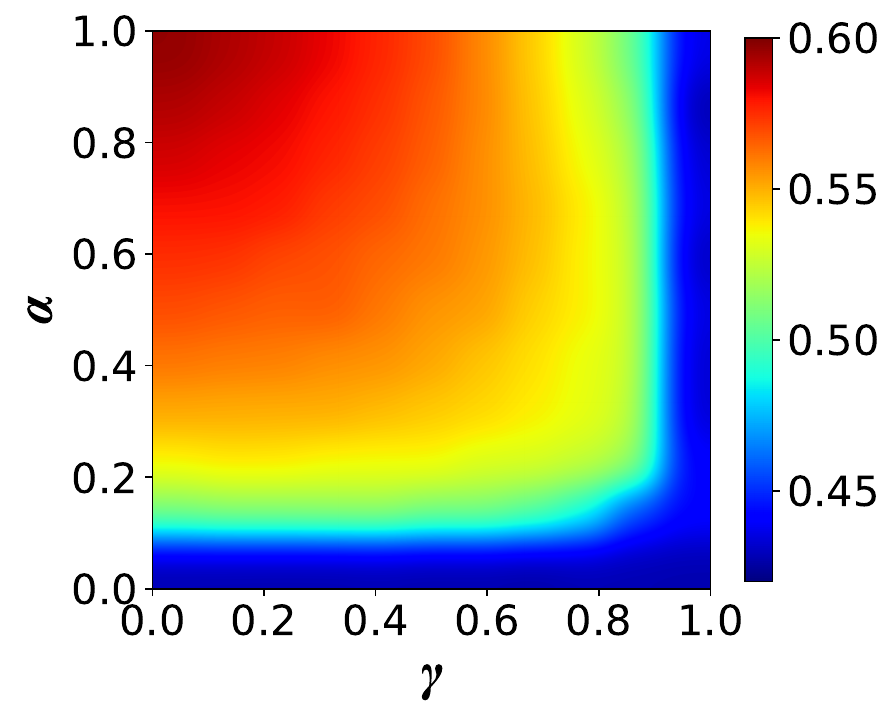}
\label{alpha_gamma_SL}}
\subfigure[WS network]{
\includegraphics[scale=0.27]{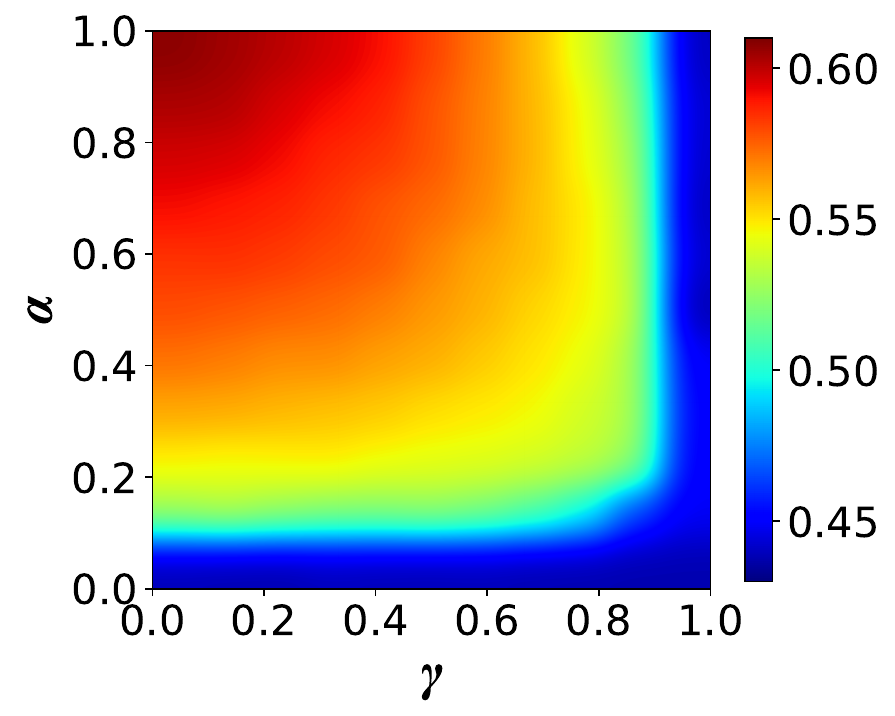}
\label{alpha_gamma_WS}}
\caption{\textbf{Heat maps of cooperation percentage with respect to learning rate and discount factor in Q-learning.} In this figure, we depict the impact of learning rate $\alpha$ and discount factor $\gamma$ on the ratio of cooperators on the SL (in panel (a)) and WS (in panel (b)) networks. We set the $x-$axis and $y-$axis in each subgraph to $\gamma$ and $\alpha$ with the range [0, 1]. These heat maps offer insights into how different combinations of learning rates and discount factors affect cooperative behavior in the two types of networks. It can be seen that a lower learning rate and a larger discount factor cause more defectors to invade the cooperators.}
\label{alpha_gamma_f}
\end{figure}
\end{center}
\vspace{-2.8\baselineskip}

\begin{center}
\begin{figure}[htbp]
\centering
\subfigure[SL network]{
\includegraphics[scale=0.27]{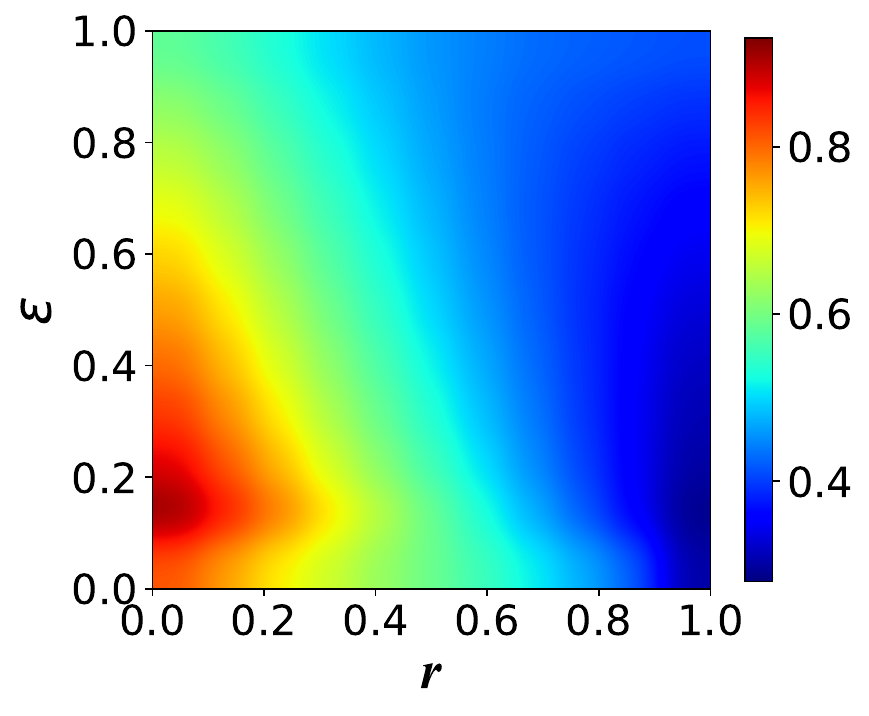}
\label{epsilon_r_SL}}
\subfigure[WS network]{
\includegraphics[scale=0.27]{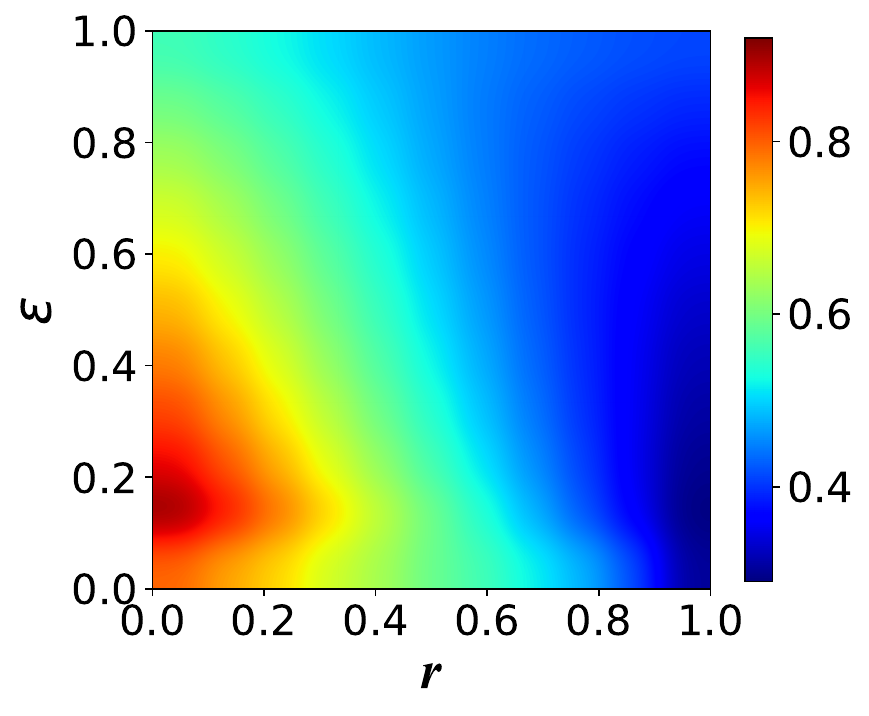}
\label{epsilon_r_WS}}
\caption{\textbf{Heat maps of cooperation ratio concerning exploration rate and payoff parameter.} In this figure, we study the influence of exploration rate $\epsilon$ and payoff parameter $r$ on the cooperative behavior within the SL (panel (a)) and WS (panel (b)) networks. The $x-$axis and $y-$axis in each subfigure represent $r$ and $\epsilon$ with a range [0, 1]. These visualizations offer insights into how variations in $\epsilon$ and $r$ affect cooperative dynamics on the SL and WS networks. We get that an appropriate exploration rate promotes cooperative behavior when $r$ is small, while increasing the exploration rate inhibits the maintenance of cooperators when $r$ is large.}
\label{epsilon_r_f}
\end{figure}
\end{center}
\vspace{-3.3\baselineskip}

It is also crucial to study the effect of learners' learning rate $\alpha$ and discount factor $\gamma$ in Eq. \ref{q-learning} on cooperative behavior. Thus, we exhibit the heat map depicting the percentage of cooperators on the SL and WS networks concerning $\alpha$ and $\gamma$ in Figs. \ref{alpha_gamma_SL} and \ref{alpha_gamma_WS}, respectively. We set the payoff parameter, the transition probabilities between profiteers and learners, the memory decay factor, and the memory length to $r = 0.5, p = q = 0.5, \beta = 0.5,$ and $M = 5$. From Figs. \ref{alpha_gamma_SL} and \ref{alpha_gamma_WS}, we can obtain that both SL and WS networks demonstrate a higher learning rate $\alpha$ leads to a larger percentage of cooperators. On the contrary, increasing the discount factor $\gamma$ results in a decrease in the cooperation frequency. This indicates that a larger learning rate and a greater tendency of learners toward immediate benefits promote the emergence and maintenance of cooperative behaviors in the network.

\begin{center}
\begin{figure*}[htbp]
\centering
\includegraphics[scale=0.09]{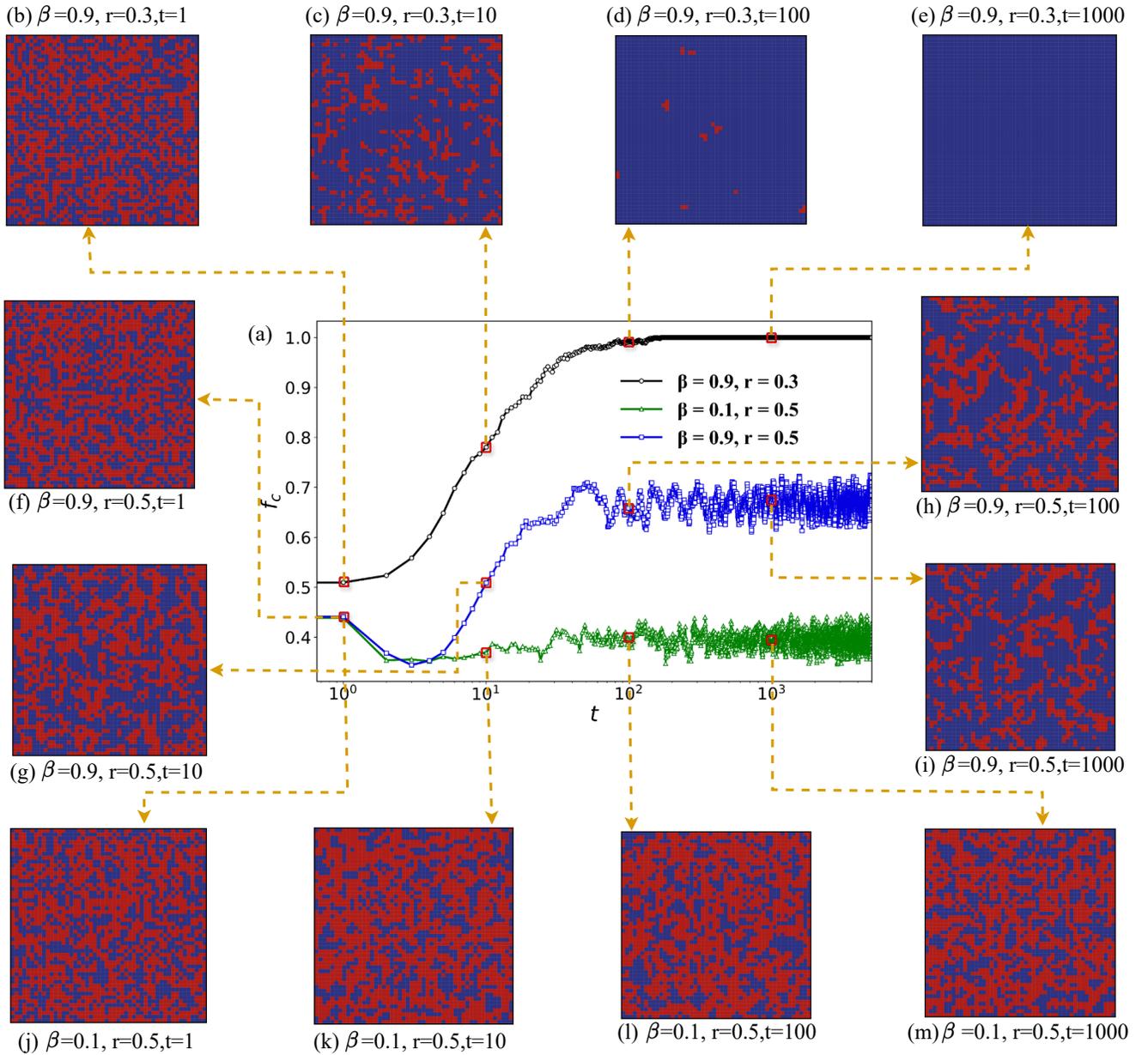}
\caption{\textbf{Evolutionary curves of cooperation frequency and corresponding snapshots under different memory decay factors and payoff parameters.} Subfigure (a) shows the evolution of cooperation frequency $f_c$ as time progresses for three distinct combinations of memory decay factor $\beta$ and payoff parameter $r$: (0.9, 0.3), (0.1, 0.5), and (0.9, 0.5). The snapshots of individual strategy distributions on the SL networks at $t = 1, 10, 100$ and 1000 are depicted in (b)-(e) for $(\beta, r)$ = (0.9, 0.3), in (f)-(i) for $(\beta, r)$ = (0.9, 0.5), and in (j)-(m) for $(\beta, r)$ = (0.1, 0.5). The memory length is fixed at $M = 10$, and the transition probabilities between learners and profiteers are set to $p = 1$ and $q = 0$. We can yield that the proportion of cooperators stabilizes around $t = 100$, suggesting that cooperators reach a stable state in the evolutionary process. Besides, cooperators resist the invasion of defectors primarily by forming clusters.}
\label{snapshot}
\end{figure*}
\end{center}
\vspace{-2.5\baselineskip}

Next, we explore the influence of exploration rate $\epsilon$ and payoff parameter $r$ on the cooperative behavior, and the results are depicted in Fig. \ref{epsilon_r_f}. With transition probabilities between profiteers and learners set to $p = 0.4, q = 0.8$, and memory mechanism parameters held constant at $\beta = 0.5$ and $M = 5$, both WS and SL networks reveal interesting insights. For relatively small values of the payoff parameter ($r<0.3$), an appropriate increase in the exploration rate fosters cooperation, but excessive exploration hampers cooperative behavior. However, when $r$ is relatively large ($r>0.3$), elevating the exploration rate consistently diminishes the prevalence of cooperators. Furthermore, we observe that an increase in the payoff parameter always reduces the fraction of cooperators, as higher $r$ values lead to greater payoffs for defectors based on Eq. \ref{Payoff matrix}, thereby incentivizing more individuals to adopt the defective strategy.

\vspace{-1.3\baselineskip}
\subsection{Snapshots of the evolution of cooperators on SL networks}
\vspace{-1.3\baselineskip}

Then, we display the evolutionary curves of the cooperation ratio under three different parameter pairs ($\beta$, $r$) and illustrate the distribution of cooperators and defectors on the SL networks at different instants from a micro perspective in Fig. \ref{snapshot}. From Fig. \ref{snapshot}(a), it is evident that the frequency of cooperators $f_c$ stabilizes around $t = 100$ in all three situations, and then stabilizes with slight fluctuations around a certain value. In addition, we observe that the cooperation ratio of $\beta = 0.9, r = 0.5$ marked by blue squares is higher than that of $\beta = 0.1, r = 0.5$ labeled by green triangles, and lower than that of $\beta = 0.9, r = 0.3$ marked by black circles, which can be visually confirmed from the snapshots depicted in Figs. \ref{snapshot}(b)-(m), where red represents the defector and blue indicates the cooperator.

Specifically, Figs. \ref{snapshot}(b), (f), and (j) all demonstrate that the network is almost evenly mixed with cooperators and defectors at $t = 1$, while the blue region for $\beta = 0.9, r = 0.3$ (Figs. \ref{snapshot}(b)-(e)) grows large as time progresses and eventually occupies the entire network, indicating that all the individuals become cooperators, and there are no defectors exist. For $\beta = 0.9, r = 0.5$ (Figs. \ref{snapshot}(f)-(i)), the blue region gradually increases and eventually stabilizes, signifying that cooperators play a dominant role in it. In contrast, the blue region gradually diminishes over time and eventually stabilizes for $\beta = 0.1, r = 0.5$ (Figs. \ref{snapshot}(j)-(m)), meaning that the defector holds a dominant role in it. Additionally, we can observe that the cooperators resist the invasion of the defectors mainly by forming clusters. These observations can be attributed to the fact that the transition probabilities between profiteers and learners are set to $p = 1$ and $q = 0$, i.e., the network is ultimately all occupied by profiteers. Moreover, we can get that increasing the memory decay factor $\beta$ promotes cooperative behaviors as indicated in Fig. \ref{mdc_T_f_profiteer}. Besides, according to Eq. \ref{Payoff matrix}, when a cooperator meets a defector, the defector's payoff increases but the cooperator's payoff decreases as the payoff parameter $r$ grows, which promotes the emergence and maintenance of defectors in the network.

\subsection{Validation of the robustness of the model}

\begin{center}
\begin{figure}[htbp]
\centering
\subfigure[SL network]{
\includegraphics[scale=0.15]{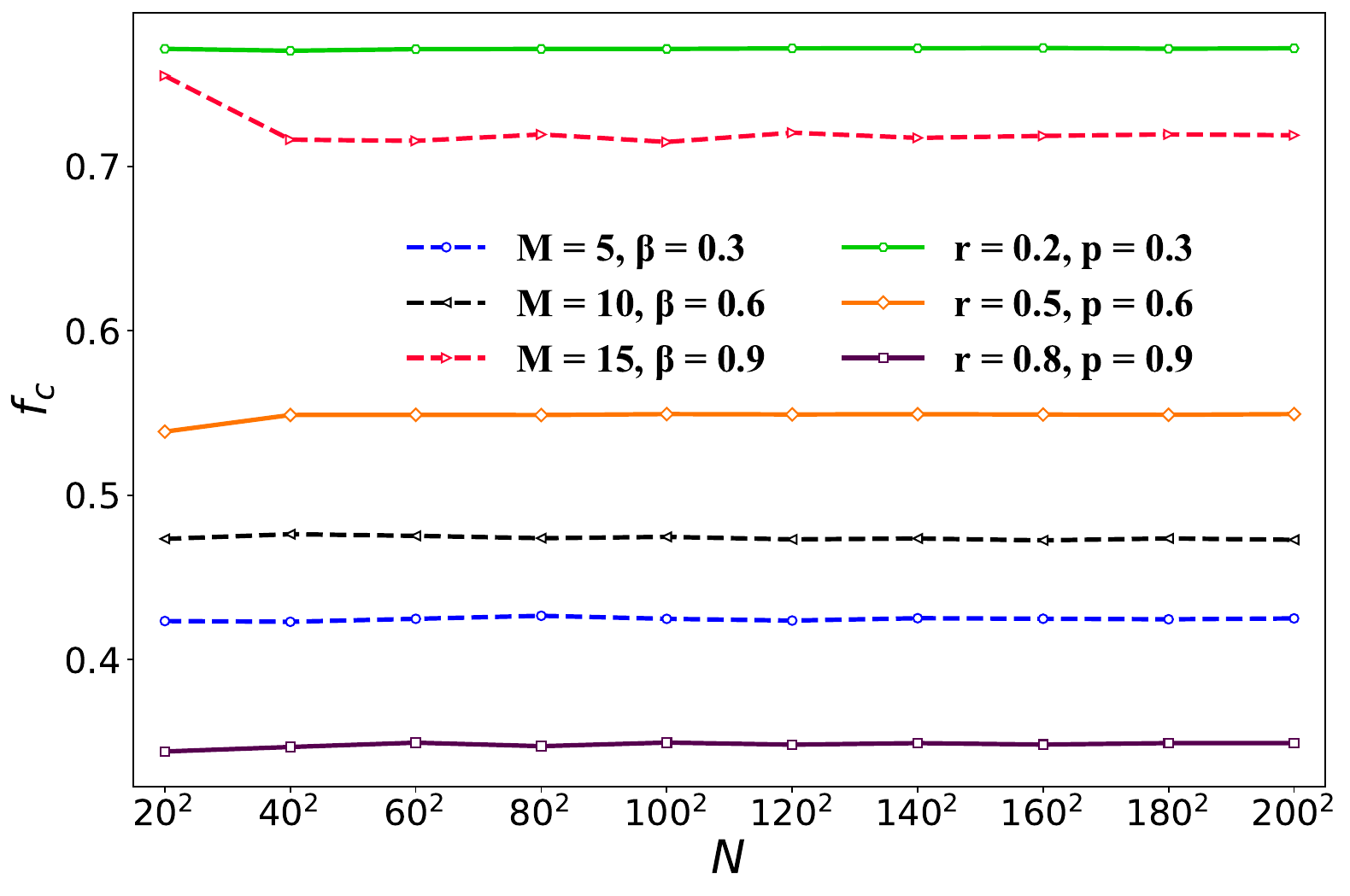}
\label{N_f_SL}}
\subfigure[WS network]{
\includegraphics[scale=0.15]{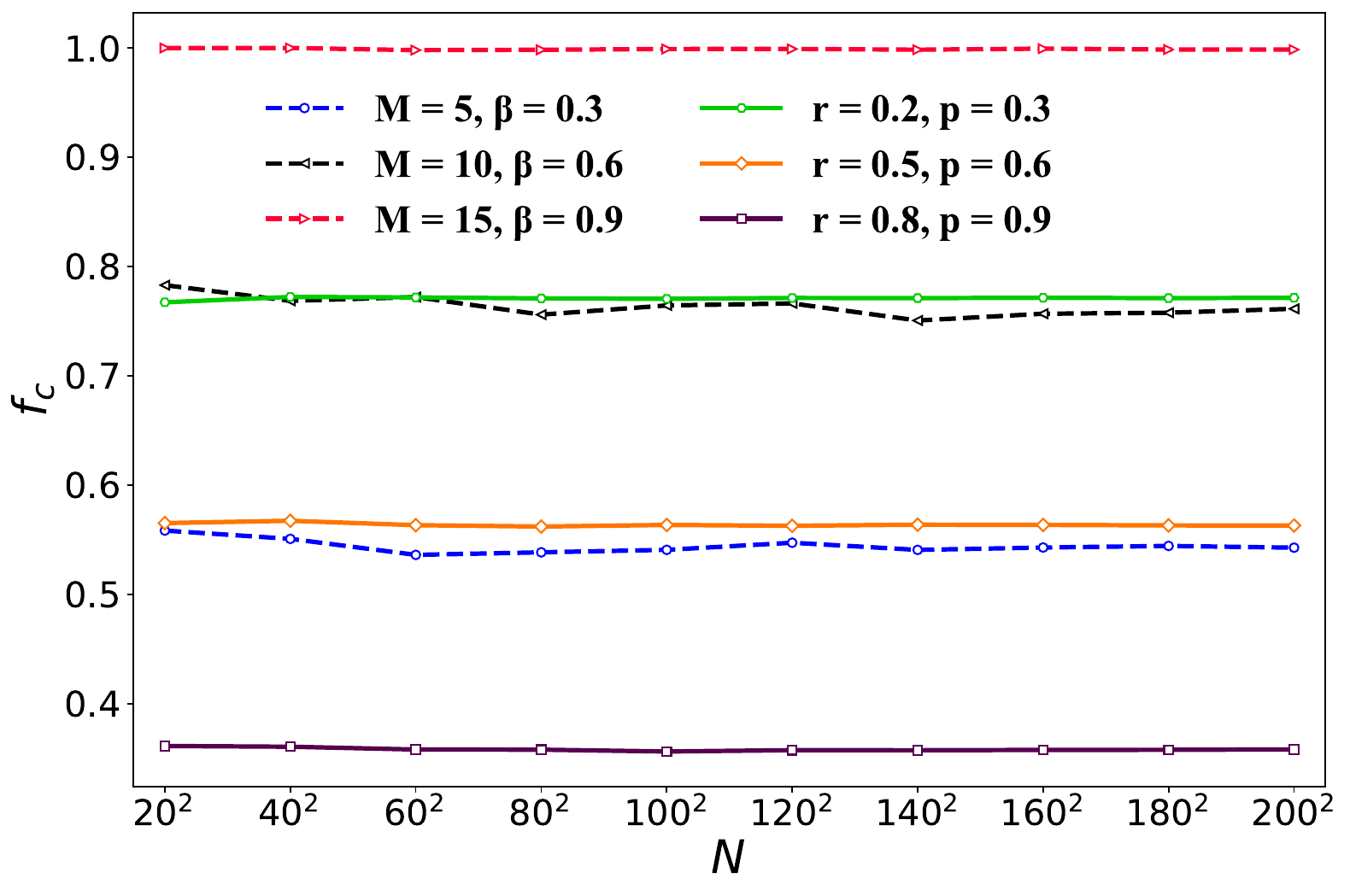}
\label{N_f_WS}}
\caption{\textbf{Variation in the proportion of cooperators with network size under different conditions.} This figure illustrates the frequency of cooperators in dependence of network size under different parameter pairs ($M$, $\beta$) and ($r$, $p$) on the SL (in subplot (a)) and WS (in subplot (b)) networks. For the ($M$, $\beta$) scenario, we set the transition probabilities between learners and profiteers to $p = 1$ and $q = 0$, payoff parameter to $r = 0.5$. In the ($r$, $p$) situation, the transition probability of the profiteer to the learner is set to $q = 0.5$, and parameters about the memory mechanism are set to $\beta = 0.6$ and $M = 10$. It is obvious that the fluctuation range and amplitude of the cooperation frequency concerning network size are minimal, which verifies the robustness of the model.}
\label{N_f}
\end{figure}
\end{center}

Note that in all of our previous simulations, the sizes of the SL and WS networks are fixed to $50 \times 50$. In this subsection, we aim to examine the evolution of cooperative behaviors on SL and WS networks with different sizes to verify the robustness of the model. We depict the variation of the cooperation frequency $f_c$ on SL and WS networks as a function of network size $N$ for three different sets of parameters ($M$, $\beta$) and three distinct groups of parameters ($r$, $p$) in Figs. \ref{N_f_SL} and \ref{N_f_WS}, respectively.

From Fig. \ref{N_f}, we obtain that both SL and WS networks exhibit almost no influence of network size on cooperative behavior under the same set of parameters, which is further demonstrated by the range and standard deviation of each curve, and the results are shown in Tabs. \ref{statistics_SL} and \ref{statistics_WS}. The reason we choose the range and standard deviation as our statistical metrics is that range indicates the extent of fluctuation in the proportion of cooperators and standard deviation quantifies the magnitude of fluctuation in the frequency of cooperators. A small range and standard deviation signify that network size has a negligible impact on the frequency of cooperators, validating the robustness of our findings across different network sizes. It can be seen that the maximum range and standard deviation of SL do not exceed 0.0402 and 0.0113, respectively, and the maximum range and standard deviation of WS are within 0.0324 and 0.0089, respectively, which signifies that the fluctuation ranges and magnitudes of $f_c$ on both networks are minimal. Notably, the fluctuation ranges and magnitudes on the WS network are even smaller than those on the SL network, consistent with the trends presented in Fig. \ref{N_f}. Furthermore, by comparing Figs. \ref{N_f_SL} and \ref{N_f_WS}, we notice that the WS network exhibits a higher percentage of cooperators for the parameter set ($M$, $\beta$) (denoted as dashed lines) compared to the SL network, while both networks show similar cooperation ratio for the parameter set ($r$, $p$) (denoted as solid lines). Therefore, through this simulation, we can conclude that the results are consistent across different network sizes for SL and WS networks, confirming the robustness of the proposed model.

\begin{center}
\begin{table*}[htbp]
\renewcommand{\arraystretch}{1.8}
\setlength{\tabcolsep}{5pt}
\begin{center}
\caption{The range and standard deviation of SL network}
\begin{tabular}{c|cccccc}
\hline
Results & $M = 5, \beta = 0.3$    & $M = 10, \beta = 0.6$    & $M = 15, \beta = 0.9$    & $r = 0.2, p = 0.3$    & $r = 0.5, p = 0.6$    & $r = 0.8, p = 0.9$ \\ \hline
Range                  & 0.0036    & 0.0037    & 0.0402   & 0.0017      & 0.0107      & 0.0054 \\
Standard deviation     & 0.0010    & 0.0011    & 0.0113   & 0.0004      & 0.0031      & 0.0016 \\
\hline
\end{tabular}
\label{statistics_SL}
\end{center}
\end{table*}
\end{center}

\begin{center}
\begin{table*}[htbp]
\renewcommand{\arraystretch}{1.8}
\setlength{\tabcolsep}{5pt}
\begin{center}
\caption{The range and standard deviation of WS network}
\begin{tabular}{c|cccccc}
\hline
Results & $M = 5, \beta = 0.3$    & $M = 10, \beta = 0.6$    & $M = 15, \beta = 0.9$    & $r = 0.2, p = 0.3$    & $r = 0.5, p = 0.6$    & $r = 0.8, p = 0.9$  \\ \hline
Range                 & 0.0222    & 0.0324    & 0.0020   & 0.0049      & 0.0053      & 0.0050  \\
Standard deviation    & 0.0062    & 0.0089    & 0.0006   & 0.0013      & 0.0014      & 0.0014 \\
\hline
\end{tabular}
\label{statistics_WS}
\end{center}
\end{table*}
\end{center}

\vspace{-4.3\baselineskip}
\section{Conclusion and outlook}
\label{Conclusion}

In this study, we investigate the evolution of cooperation on SL and WS networks with dynamic interactions between learners and profiteers, considering the category of individuals as a homogeneous discrete-time Markov chain with two states. Different categories of individuals update their strategies according to different rules, where learners adopt Q-learning while profiteers follow the Fermi rule. Additionally, we introduce the memory decay factor and memory length to enable individuals to compute payoffs based on a broader historical context rather than solely relying on the current round of the game. In the simulation, we begin with plotting the evolutionary curve and statistical distribution of learners and verify the theoretical analyses through various numerical features. Subsequently, we perform numerous simulations about the effect of the proposed model on cooperative behavior on the SL and WS networks. We find that dynamic interactions between profiteers and learners promote cooperation, increasing learning rate and decreasing discount factor in Q-learning increase the proportion of network cooperators, and memory mechanisms enhance the emergence of cooperation in pure profiteer groups. Then, we perform snapshots of the evolution of cooperators on SL networks and focus on the formation and evolution of the cooperation clusters from a micro perspective. We discover that cooperators resist the invasion of defectors mainly by forming cluster structures, with smaller payoff parameters and larger memory decay coefficients leading to larger clusters. Furthermore, our simulations on networks of varying sizes demonstrate the robustness of the model, revealing that network size has a negligible impact on the cooperation ratio on both SL and WS networks.

Based on our work, there are some extensions to further exploration in spatial evolutionary games. For example, we mainly focus on the dynamic interactions between profiteers and learners on static networks, while there actually exist structured networks that undergo dynamic changes over time \cite{holme2012temporal, li2020evolution}. These networks feature evolving interactions among individuals, presenting an intriguing opportunity to extend our model to investigate temporal networks. In addition to examining complete interactions, more and more scholars have recently turned their attention to stochastic and incomplete games \cite{li2021evolution, li2022impact, wang2021evolution}, which is also worth considering in our future research endeavors. Moreover, some previous studies have pointed out that real-world scenarios often involve conformist behaviors \cite{pi2022evolutionary, szolnoki2015conformity}. Therefore, future research could explore the introduction of conformists and assess the impact of the three categories of individual dynamic interactions on the maintenance and evolution of cooperative behavior.

\section*{Acknowledgment}

This work was supported in part by the National Natural Science Foundation of China (NSFC) under Grant No. 62206230 and No. 12271083 and in part by the Natural Science Foundation of Sichuan Province under Grant No. 2022NSFSC0501.




\section*{References}


\end{document}